 \documentclass[prb,twocolumn,showpacs,showkeys,floatfix,superscriptaddress]{revtex4}

\usepackage{graphicx}
\usepackage{amssymb,amsmath}
\usepackage{dcolumn}
\usepackage{bm}
\usepackage{color}
\usepackage{tabularx}

\begin{document}

\title{Electronic structure and thermoelectric properties of 
       $ {\bf CuRh_{1-x} Mg_xO_2} $}

\author{Antoine Maignan}
\affiliation{Laboratoire CRISMAT, UMR CNRS-ENSICAEN(ISMRA) 6508, 
             and IRMA, FR3095, Caen, France}
\author{Volker Eyert}
\affiliation{Laboratoire CRISMAT, UMR CNRS-ENSICAEN(ISMRA) 6508, 
             and IRMA, FR3095, Caen, France}
\affiliation{Center for Electronic Correlations and Magnetism, 
             Institut f\"ur Physik, Universit\"at Augsburg, 
             86135 Augsburg, Germany}
\author{Christine Martin}
\affiliation{Laboratoire CRISMAT, UMR CNRS-ENSICAEN(ISMRA) 6508, 
             and IRMA, FR3095, Caen, France}
\author{Stefan Kremer}
\affiliation{Laboratoire CRISMAT, UMR CNRS-ENSICAEN(ISMRA) 6508, 
             and IRMA, FR3095, Caen, France}
\affiliation{Institut f\"ur Theorie der  Kondensierten Materie,
             Universit\"at Karlsruhe, 76128 Karlsruhe, Germany}
\author{Raymond   Fr\'esard}
\affiliation{Laboratoire CRISMAT, UMR CNRS-ENSICAEN(ISMRA) 6508, 
             and IRMA, FR3095, Caen, France}
\author{Denis Pelloquin}
\affiliation{Laboratoire CRISMAT, UMR CNRS-ENSICAEN(ISMRA) 6508, 
             and IRMA, FR3095, Caen, France}

\date{\today}

\begin{abstract}
Electronic structure calculations using the augmented spherical 
wave method have been performed for $ {\rm CuRhO_2} $. For this 
semiconductor crystallizing in the delafossite structure, it is found 
that the valence band maximum is mainly due to the $ 4d $ $ t_{2g} $ 
orbitals of $ {\rm Rh^{3+}} $. The structural characterizations of 
$ {\rm CuRh_{1-x} Mg_xO_2 } $ show a broad range of $ {\rm Mg^{2+}} $ 
substitution for $ {\rm Rh^{3+}} $ in this series, up to about 12\%. 
Measurements of the resistivity and thermopower of the doped systems 
show a Fermi liquid-like behavior for temperatures up to about 1000\,K, 
resulting in a large weakly temperature dependent power factor. The 
thermopower is discussed both within the Boltzmann equation approach 
as based on the electronic structure calculations and the temperature
independent correlation functions ratio approximation as based on the Kubo
formalism.  
\end{abstract}

\pacs{71.20.-b,  
      72.15.Eb,  
      73.90.+f 
      }

\keywords{electronic structure, low-dimensional compounds}

\maketitle

\section{Introduction}
\label{sec:intro}

The search for new thermoelectric materials in order to convert waste-heat
into electricity has motivated numerous studies on transition-metal
oxides. One advantage of these materials over several others lies in 
their ability to be used at elevated temperatures in air. This opens 
the route to target systems releasing heat at temperatures as high as 
1000\,K. Among the studied p-type thermoelectric oxides, the layered 
ones such as $ {\rm Na_xCoO_2} $, misfit cobaltites, or, more recently,  
the delafossites, all with structures containing $ {\rm CdI_2} $-type 
layers, have been particularly investigated according to the richness 
of their physical properties. For instance, the thermoelectric performance 
of several $ {\rm AMO_2} $ delafossites have been measured, leading to 
the following dimensionless figures of merit $ ZT $ of 0.04 at 800\,K 
($ {\rm CuCr_{0.97}Mg_{0.03}O_2} $), \cite{Zcucrmg} 0.14 at 1100\,K 
($ {\rm Cu Fe_{0.99}Ni_{0.01}O_2} $), \cite{Zcufeni} and 0.15 at 1000\,K 
($ {\rm Cu Rh_{0.90}Mg_{0.10}O_2} $). \cite{kuriyama06} Their crystal 
structures can all be described as a delafossite-type, in which 
$ {\rm MO_2} $-layers of edge sharing $ {\rm MO_6} $ octahedra alternate 
along the c-axis with layers of monovalent $ {\rm Cu^+} $ cations, 
the latter exhibiting a dumbbell O--Cu--O coordination. 

Although the thermoelectric properties of the Cu--based delafossites 
have been measured, their origin remains a subject of controversy. 
In the $ {\rm CuCr_{1-x}Mg_xO_2} $ system, the measurements have been 
interpreted by considering different active layers for the electrical 
transport -- either in the $ {\rm Cu} $ or $ {\rm CrO_2} $ layers -- 
whereas in a recent report on $ {\rm CuRhO_2} $, the electrical 
conductivities of both $ {\rm Cu} $ and $ {\rm RhO_2} $ layers have 
been proposed to be comparable at 300\,K. \cite{shibasaki06} 

In order to shed light on the respective role of the layers on the 
transport properties in delafossites, electronic structure 
calculations have proven to be useful.
\cite{galakhov97,seshadri98,ong07,singh07,singh08,pdcoo2,cufeo2,cucro2,usui09}  
In the $ {\rm ACoO_2} $ delafossite with A = Pt or Pd, these calculations 
have demonstrated that their low resistivities 
($ \sim 5\,\mu\Omega~{\rm cm} $ at room
temperature\cite{Tanaka96,dupont71}) come almost exclusively from the in-plane  
$ d $ orbitals of the $ {\rm A^+} $ cations. \cite{seshadri98,pdcoo2} 
For $ {\rm CuMO_2} $, on the opposite, it is found that the $ t_{2g} $ 
states of the $ {\rm M=Cr^{3+}} $ cations provide the most important 
contribution at the valence band maximum, with spin polarization 
supported by the experimental evidences for negative magnetoresistance 
and magnetothermopower.\cite{cucro2} However, as has been recently 
reported for $ {\rm CuYO_2} $,\cite{singh08} where the $4d^0$ stable 
electronic configuration of $ {\rm Y^{3+}} $ precludes any 
participation of Y to the transport, there exists some cases where 
the copper cations are contributing to the charge delocalization. However,
this delafossite belongs to those characterized by a large M cation
($r_{\rm Y^{3+}} = 0.090 $\,nm versus $ r_{\rm Cr^{3+}} = 0.0615 $\,nm) 
favoring the incorporation of extra $ {\rm O^{2-}} $ anions into the copper 
layer so that in that case the physics might be different. In that respect, the
physics of the oxygen stoichiometric $ {\rm CuMO_2} $ subclass of delafossites
is of interest, especially if we consider the possibility to control the
magnetism at the M-site. This is outlined by the multiferroic behavior
exhibited by the $ {\rm CuMO_2} $ delafossites for $ {\rm M=Cr^{3+}} $ or 
$ {\rm Fe^{3+}} $, which are both showing electric polarization induced by
incommensurate antiferromagnetism according to the  $\mathit{S}=\frac{3}{2} $ and 
$\mathit{S}=\frac{5}{2} $  high-spin M cations, 
respectively.\cite{doumerc86,kimura06,seki08} For these delafossites, 
the complex magnetism resulting from the frustrated nature of the 
$ {\rm MO_2} $ network is indeed thought to be responsible for the
magnetic field induced electric polarization.

Such multiferroic behavior is in marked contrast to the metal-type
behavior reported for the $ {\rm Mg^{2+}} $ substituted rhodate 
$ {\rm CuRh_{1-x}Mg_xO_2} $. The $ 4d^5 $/$ 4d^6 $ electronic
configurations of low-spin $ {\rm Rh^{4+}/Rh^{3+}} $ in this series 
provides a unique opportunity to study the electronic groundstate of a 
delafossite without extra contributions generated by the large magnetic 
moments of $ {\rm Cr^{4+}} $/$ {\rm Cr^{3+}} $ ($ \mathit{S} = 1/\mathit{S} =\frac{3}{2} $) or 
$ {\rm Fe^{4+}} $/$ {\rm Fe^{3+}} $ ($\mathit{S} = 2/\mathit{S} =\frac{5}{2} $) cations. 
However, literature data for $ {\rm CuRh_{1-x}Mg_xO_2} $ show some 
variation. This is best illustrated by the different values reported 
for the room temperature Seebeck coefficient $S$ of 
$ {\rm CuRh_{0.90}Mg_{0.10}O_2} $, which are 
$ S_{\rm 300K} = 130\,{\rm \mu V K^{-1}} $,
and  $ S_{\rm 300K} =  70\,{\rm \mu V K^{-1}} $ as given in Refs.\ 
\onlinecite{kuriyama06} and \onlinecite{shibasaki06}, respectively.\cite{kuriyamarem} 
Among the possible reasons explaining this discrepancy is the 
uncertainty of the Mg for Rh substitution controlling the concentration 
of $ {\rm Rh^{4+}} $ holes in the $ {\rm Rh^{3+}} $ matrix. In order to 
check for the transport mechanism and for the substitution effectiveness, 
electronic structure calculations have been performed. In the 
present paper, these results are discussed in comparison to the 
experimental data. The latter are obtained on polycrystalline samples, 
for which the delafossite structure was verified by a combined X-ray 
and electron diffraction study. Taken together with the cation analysis, 
these characterizations also demonstrate the existence of a 
$ {\rm Mg^{2+}} $ solubility limit much larger than in 
$ {\rm CuCr_{1-x}Mg_xO_2} $. The electrical resistivity and Seebeck
coefficient measurements show that the substitution of Mg for Rh in 
$ {\rm CuRhO_2} $ allows to progressively induce an insulator to metal
transition accompanied by a gradual decrease of the thermopower.  

Contrasting previous studies, these results are found to be compatible
with electronic structure calculations with a lack of significant
contribution of copper, the most important contribution at the Fermi level
coming from the $ 4d $ orbitals of the Rh cations. The lack of significant
magnetic contribution, probed by magnetic susceptibility, responsible for 
the low temperature increase in $ {\rm  CuCrO_2} $ and $ {\rm CuFeO_2} $, 
allows for the study of the metallic state from 2.5\,K to 1000\,K. An unusual 
$T^2$ regime is evidenced over a broad temperature range, which, when 
combined with the $ T $ dependence of the Seebeck coefficient, leads to 
remarkably $T$ independent power factors, 
$ {\rm PF} \equiv \frac{S^2}{\rho} $.

\section{Methodology}

\subsection{Electronic structure calculations: theoretical method}
\label{sec:thmet}

The calculations are based on density-functional theory and the 
generalized gradient approximation (GGA). \cite{perdew96a} They 
were performed using the scalar-relativistic implementation of 
the augmented spherical wave 
(ASW) method (see Refs.\ \onlinecite{wkg,aswrev,aswbook} and references 
therein). In the ASW method, the wave function is expanded in atom-centered
augmented spherical waves, which are Hankel functions and numerical
solutions of Schr\"odinger's equation, respectively, outside and inside
the so-called augmentation spheres. In order to optimize the basis set,
additional augmented spherical waves were placed at carefully selected
interstitial sites. The choice of these sites as well as the augmentation
radii were automatically determined using the sphere-geometry optimization
algorithm.\cite{sgo} Self-consistency was achieved by a highly efficient
algorithm for convergence acceleration.\cite{mixpap} The Brillouin zone
integrations for the self-consistent field calculations were performed 
using the linear tetrahedron method with up to 1156 {\bf k}-points within 
the irreducible wedge of the rhombohedral Brillouin 
zone,\cite{bloechl94,aswbook} whereas the calculation of the density 
of states and the thermopower is based on 19871 {\bf k}-points. 

In the present work, a new full-potential version of the ASW method 
was employed.\cite{fpasw}
In this version, the electron density and related quantities are given
by spherical-harmonics expansions inside the muffin-tin spheres.
In the remaining interstitial region, a representation in terms of
atom-centered Hankel functions is used.\cite{msm88} However, in
contrast to previous related implementations, we here get away without
needing a so-called multiple-$ \kappa $ basis set, which fact allows to 
investigate rather large systems with a minimal effort.

\subsection{Ceramic samples preparation and characterization}

The polycrystalline samples of the $ {\rm CuRh_{1-x}Mg_xO_2} $ series 
have been prepared by solid state reaction in air. Bars of typical 
size $ 2 \times 2 \times 10 $\,mm were prepared by mixing stoichiometric 
amounts of the $ {\rm Cu_2O} $, $ {\rm Rh_2O_3} $ and MgO precursors, 
which were then pressed. The electron
diffraction study was carried out with a JEOL 2010CX transmission 
electron microscope. The sample preparation is made by crushing in 
butanol some bar pieces, the corresponding microcrystals being 
afterwards deposited on Ni grids. The purity of the 
obtained black bars was checked by X-ray powder diffraction using a 
Panalytical X-pert Pro and a Brucker diffractometers. The data were 
analyzed by using the Fullprof suite.\cite{Rodriguez93} 

\begin{figure}[tb]
\centering
\includegraphics[width=\columnwidth,clip]{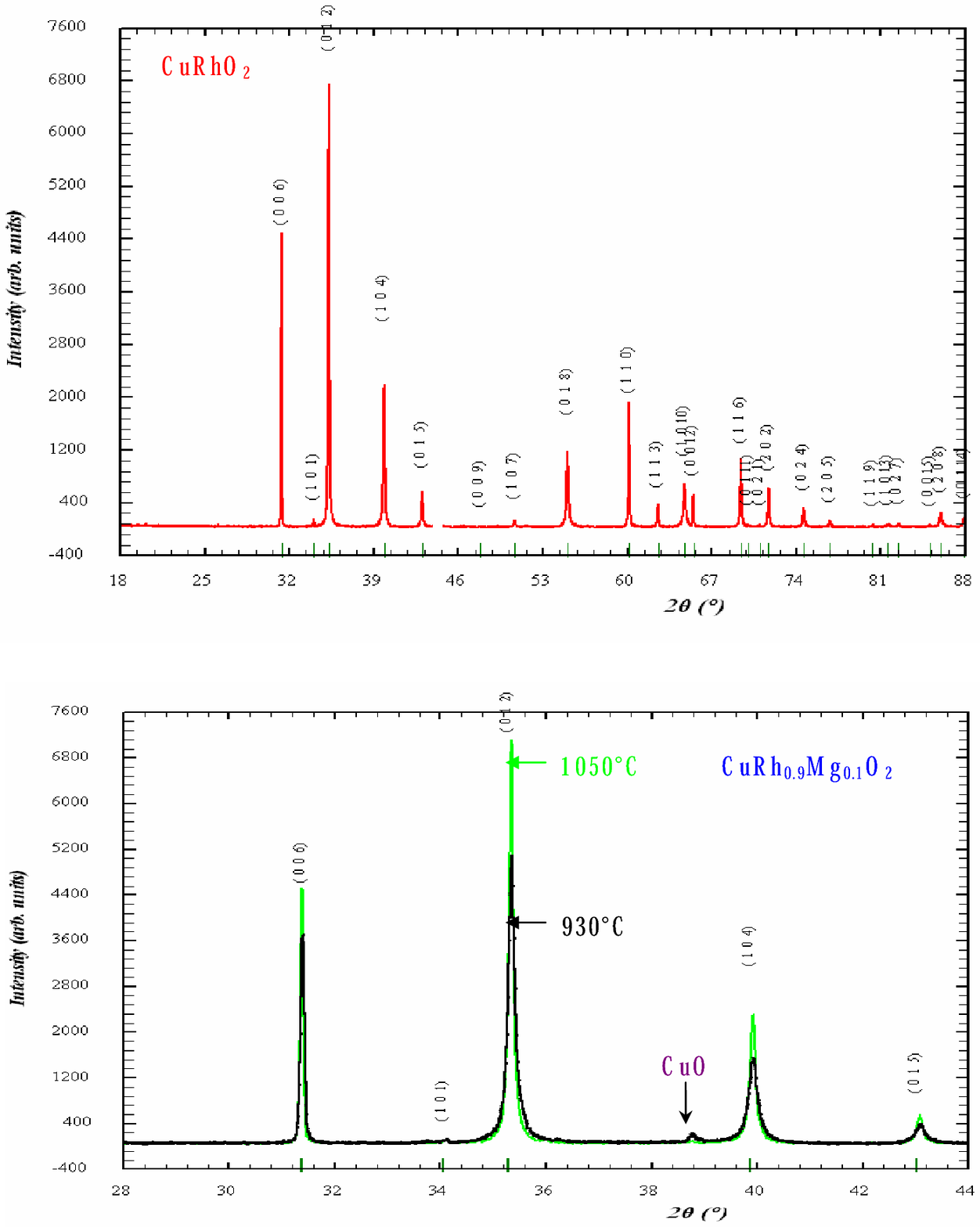}
\includegraphics[width=\columnwidth,clip]{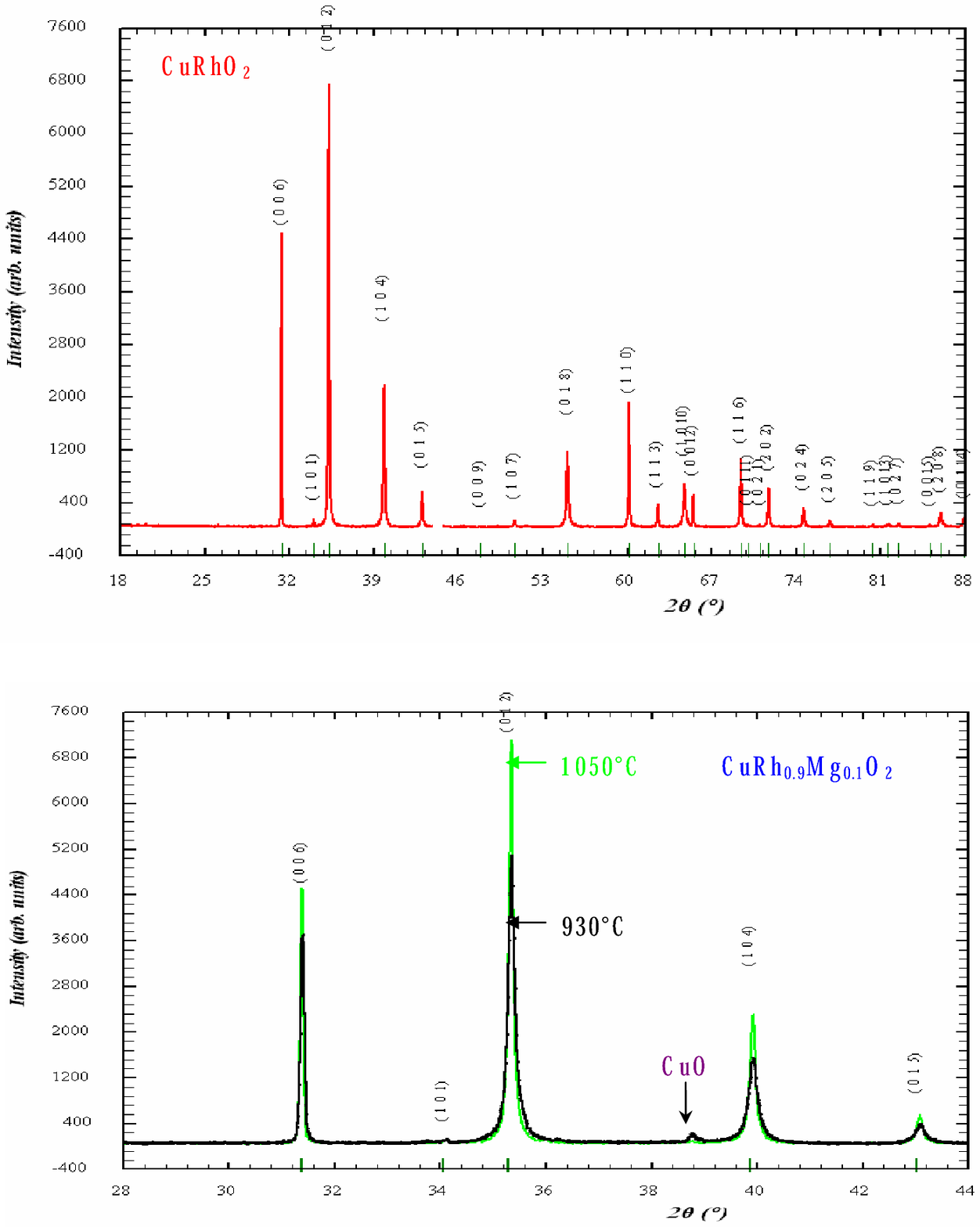}
\caption{(Color online) 
         Top: X-ray patterns of two $ {\rm Cu Rh_{0.90}Mg_{0.10}O_2} $ 
         samples synthesized at $930^\circ$C and $1050^\circ$C, respectively, 
         with indexation in the $R\bar{3}m$ space group. 
         Bottom: X-ray diffraction patterns at room temperature; the vertical
         bars correspond to the locations of the diffraction peaks in the 
         $R\bar{3}m$ space group (with $a=3.0741(1)$\,{\AA}, and 
         $c=17.0952(3) $\,{\AA}). The missing data in the 44-44.8$^\circ$ 
         region ($2\theta$) have been intentionally eliminated due to a 
         small peak coming from the sample holder. } 
\label{fig:struc}
\end{figure}

As two different calcination temperatures have been used in the literature,
two attempts were made at either $930^\circ$C or $1050^\circ$C for a duration
of twelve hours. As shown in Fig.~\ref{fig:struc}a for 
$ {\rm CuRh_{0.9}Mg_{0.1}O_2} $ ($ {\rm x=0.1} $), one small extra peak
identified as CuO appears (see the vertical arrow in
Fig.~\ref{fig:struc}a). The weight of the latter depends on the synthesis 
temperature. Indeed, when increasing it from $930^\circ$C to $1050^\circ$C,
the weight of this peak decreases. Furthermore, as these patterns were
recorded in the same conditions, (amount of powder and acquisition time), this
comparison shows the poor crystallinity of the $930^\circ$C prepared
sample. This result agrees with the less dispersed cation contents of the
$1050^\circ$C prepared sample as probed by energy dispersive X-ray
spectroscopy (EDS) analysis performed within the transmission electron
microscope. Such conclusions are consistent with the thermodynamics of the
Cu-Rh-O ternary diagram, which showed that $ {\rm CuRhO_2} $ synthesized  
below $985^\circ$C in an oxygen pressure of 0.1\,MPa is unstable. 
\cite{Jacob99} Thus, the sample series corresponding 
to $ {\rm x=0} $, 0.01, 0.04, 0.10, 0.15, and 0.30 in 
$ {\rm CuRh_{1-x}Mg_xO_2} $ has been calcined in air
at $1050^\circ$C for a duration of twelve hours.

The patterns of the $ {\rm CuRh_{1-x}Mg_xO_2} $ samples are all refined
in the $R\bar{3}m$  space group, usually reported for delafossite 
compounds at room temperature, as shown in Fig.~\ref{fig:struc}. 
A small decrease of both $a$ and $c$ lattice parameters is observed when x 
increases up to 15\% (corresponding to a decrease of $\sim$ 0.6\% of the 
cell volume). From $ {\rm x=0.10} $ on, one small extra peak appears that can 
be identified as CuO. 

In order to test the maximum content of $ {\rm Mg^{2+}} $ substituted 
for Rh cations, EDS analyses coupled to electron diffraction have been 
made first for the compound with the highest experimental Mg doping 
($ {\rm x=0.3} $). This analytical study demonstrates that a maximum of 
12\% $ {\rm Mg^{2+}} $ can be substituted at the Rh site in 
$ {\rm  CuRhO_2} $, while the $ {\rm Cu_2MgO_3} $ oxide is detected as
a secondary phase. This is consistent with the observation from X-ray
diffraction of this impurity for ${\rm x > 0.10}$. For that reason, our
measurements of transport properties are restricted to compounds 
corresponding to $ {\rm x} \leq 0.10$, the $ {\rm x}=0.10$ composition 
already containing a very low amount of CuO as impurity. We emphasize that
this structural study cannot be reconciled with previous data by Shibasaki et
al.\cite{shibasaki06} given for compounds nominally containing up to $20\%$ of
$ {\rm Mg^{2+}} $.

The low temperature ($ T<320 $\,K) electrical resistivity ($\rho$) 
and Seebeck coefficient ($S$) were measured by using a Quantum Design 
cryostat. The four-probe and the steady-state techniques were used for 
the former and the latter, respectively, indium electrical contacts 
having been deposited with ultrasons. For the high temperature $\rho$ 
and $S$ measurements ($T>300\,K$) a Ulvac-Zem~3 system was used. The 
magnetic susceptibility was measured by using a dc SQUID magnetometer 
(zero field-cooling, $\mu_0H= 0.3 $\,T).

\section{Results}

\subsection{Calculations: Important role of the Rh $ 4d $ orbitals}
\label{sec:calc}

The calculations were based on the crystal structure data by Oswald 
{\em et al.},\cite{oswald89} who determined the lattice constants 
as $ a = 3.08 $\,{\AA} and $ c = 17.09 $\,\AA. However, these authors 
did not measure the internal oxygen parameter. For this reason, we 
performed an energy minimization, leading to a value of 
$ z_{\rm O} = 0.10717 $, which was used in all subsequent calculations. 

\begin{figure}[htb]
\centering
\includegraphics[width=0.8\columnwidth]{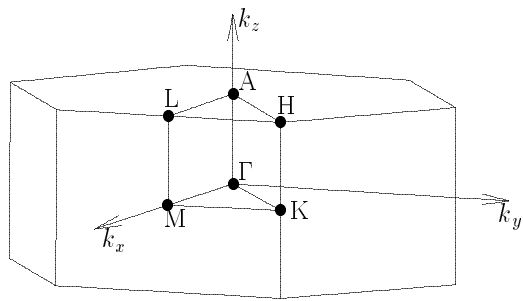}
\caption{First Brillouin zone of the hexagonal lattice.}
\label{fig2}
\end{figure}

\begin{figure}[b]
\centering
\includegraphics[width=\columnwidth,clip]{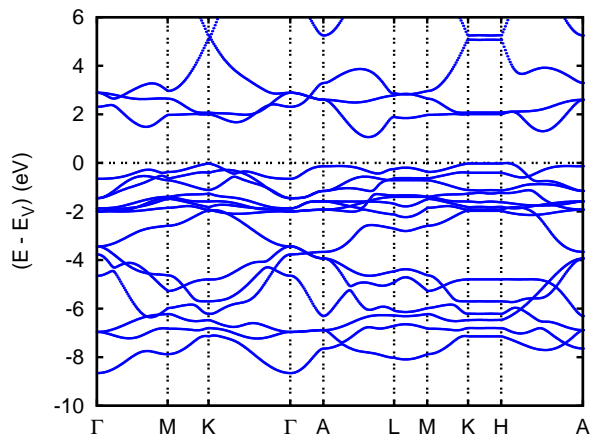}
\caption{(Color online) Electronic bands of $ {\rm CuRhO_2} $.}
\label{fig3}
\end{figure}

The electronic bands along selected high-symmetry lines of the first
Brillouin zone of the hexagonal lattice, Fig.~\ref{fig2}, are displayed in
Fig.~\ref{fig3}. 

\begin{figure}[tb]
\centering
\includegraphics[width=\columnwidth,clip]{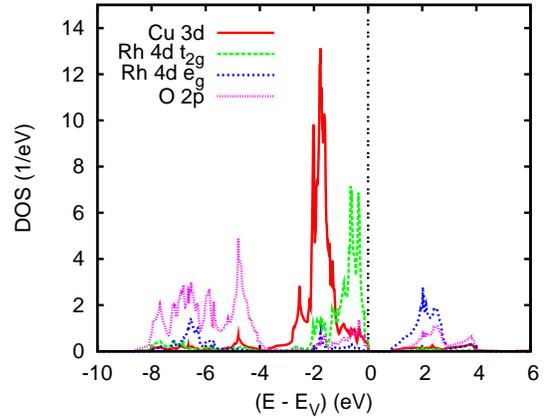}
\caption{(Color online) Partial densities of states (DOS) of 
          $ {\rm CuRhO_2} $. Selection of the Rh $ 4d $ orbitals 
          is relative to the local rotated reference frame, 
          see text.}
\label{fig4}
\end{figure}

The corresponding partial densities of states (DOS) are shown in
Fig.~\ref{fig4}.
While the lower part of the spectrum is dominated by O $ 2p $ states, 
the transition metal $ d $ states lead to rather sharp peaks in the 
interval from $ -4 $ to $ +4 $\,eV. In particular, the $ t_{2g} $ 
and $ e_g $ manifolds of the Rh $ 4d $ states as resulting from the 
octahedral coordination are recognized. This representation of the 
partial DOS used a local rotated coordinate system with the Cartesian 
axes pointing towards the oxygen atoms. $ \sigma $-type overlap of the 
O $ 2p $ states with the Rh $ 4d $ $ e_g $ orbitals leads to the 
contribution of the latter between $ - 7 $ and $ -6 $\,eV. In contrast, 
due to the much weaker $ \pi $-type overlap of the O $ 2p $ states with 
the $ t_{2g} $ orbitals, these states give rise to sharp peaks in the 
interval from $ -2.2 $\,eV to the valence band maximum. The $ t_{2g} $ 
manifold is separated by an optical band gap of $ \approx 0.75 $\,eV 
from the empty $ e_g $ states and thus leads to a Rh $ d^6 $ state. 
The Cu $ 3d $ states are essentially limited to the interval from 
$ -4 $\,eV to the valence band maximum and thus Cu can be assigned a 
monovalent $ d^{10} $ configuration in close analogy with the 
experimental findings. In passing, we mention the finite dispersion 
of the electronic bands parallel to $ \Gamma $-A, which points to a 
considerable three-dimensionality arising from the coupling between 
the layers. Yet, this overall behavior of the dispersion perpendicular 
to the $a-b$ plane is contrasted by the barely noticeable dispersion 
particularly along the line K-H. This has been also observed for 
other delafossite materials.\cite{pdcoo2,cucro2} 

\begin{figure}[htb]
\centering
\includegraphics[width=\columnwidth,clip]{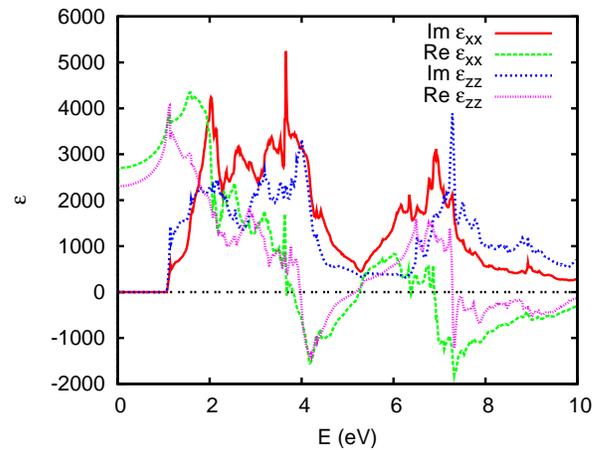}
\caption{(Color online) Dielectric function of $ {\rm CuRhO_2} $.}
\label{fig7}
\end{figure}

Further insight about the electronic properties of $ {\rm CuRhO_2} $ can be
gained by analyzing the real and imaginary parts of the dielectric function as
calculated within linear-response (see Ref.\ \onlinecite{aswbook} for more
details). As is evident from Fig.~\ref{fig7}, 
the asymmetry between the in-plane and out-of-plane directions is not 
reflected in the absorption gap following from the imaginary parts of 
the dielectric function. In fact the gap is very close to 0.75\,eV in 
all three directions, which is most likely to exceed the Hund's rule 
coupling. Therefore, the low-spin $ 4d^6 $ configuration of 
$ {\rm Rh^{3+}} $ is expected to be the ground state. This is
consistent with earlier findings by Singh for $ {\rm CuCoO_2} $, where  
the Co ions adopt the low-spin $ 3d^6 $ configuration.\cite{singh07}

Finally, we have calculated the thermopower using the framework of 
Boltzmann theory.\cite{allen96} The transport properties are 
expressed in terms of the Onsager transport coefficients, 
\begin{equation}
L_{\lambda \lambda'}^{(n)} 
  =   \frac{ 1 }{ T } \int_{-\infty}^{+\infty} d E \, 
      \left( - \frac{ \partial f ( E ) }{ \partial E } \right)
      \Xi_{\lambda \lambda'} ( E ) 
      \left( E - \mu \right)^n  
                        \;, 
\label{eq:aswft16} 
\end{equation}
where $ \left( - \frac{ \partial f ( E ) }{ \partial E } \right) $ is 
the negative derivative of the Fermi function and 
\begin{equation}
\Xi_{\lambda \lambda'} ( E ) 
  =   \frac{ 1 }{ \Omega_c } 
      \sum_{\bf k} \sum_{n} 
      v_{ {\bf k} n}^{\lambda  }
      v_{ {\bf k} n}^{\lambda' }  
      \tau_{ {\bf k} n}
      \delta ( E - \varepsilon_{ {\bf k} n} ) 
\label{eq:aswft13}
\end{equation}
denotes the so-called transport distribution.\cite{scheidemantel03,madsen06} 
Here, $ \Omega_c $ is the volume of the unit cell, 
$ v_{ {\bf k} n}^{\lambda  } $ a Cartesian component of the group velocity 
of the $ n $'th band, and $ \tau_{ {\bf k} n} $ is the relaxation time. 
While the electrical conductivity and the thermal conductivity (at zero 
electric field) are given directly by (\ref{eq:aswft16})  for $ n = 0 $ 
and $ n = 2 $, respectively, the thermopower is calculated from the matrix 
equation 
\begin{equation}
S_{\lambda \lambda'}  
  =   \frac{1}{eT} 
      \left( \left[ L^{(0)} \right]^{-1} L^{(1)} 
      \right)_{\lambda \lambda'} 
                        \;,
\label{eq:aswft17}
\end{equation}
where $ e $ is the (negative) electronic charge. 
Following standard practice we assume that the relaxation time does not 
depend on $ {\bf k} $ and band index, in which case $ \tau $ 
cancels from the Seebeck coefficient. The implementation of the above 
formulation was done along similar lines as those previously proposed 
by Scheidemantel {\em et al.}\ as well as by Madsen and 
Singh.\cite{scheidemantel03,madsen06} The implementation was tested 
against the recent results by Singh for $ {\rm CuCoO_2} $ and 
$ {\rm YCuO_2} $ and very good agreement was found. \cite{singh07,singh08} 

\begin{figure}[htb]
\centering
\includegraphics[width=\columnwidth,clip]{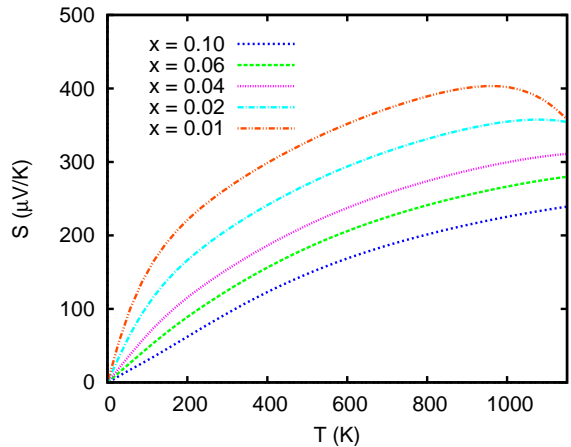}
\caption{(Color online) Thermopower $ S_{xx} $  of $ {\rm CuRhO_2} $ 
         for different hole doping levels.} 
\label{fig8}
\end{figure}

The $ xx $-components of the thermopower as calculated for different 
doping levels are displayed in Fig.~\ref{fig8};  
the calculated $ zz $-components are about 10-20\% larger. According to 
these findings, the thermopower strongly decreases with increased hole 
doping. In addition, it shows an almost linear dependence on temperature 
especially in the intermediate-temperature range with the downturn at 
low doping and high temperatures reflecting excitations across the 
optical band gap. Worth mentioning are the rather high values for the 
lower doping levels down to $ \approx 100 $\,K and the pronounced 
drop below this temperature. 

In passing, we mention that apart from systematically slightly smaller 
values our results are in perfect agreement with the calculations of 
Usui {\em et al.},\cite{usui09} who likewise used the Boltzmann 
equation approach and who in turn obtained almost perfect agreement 
with the experimental data by Kuriyama {\em et al.}.\cite{kuriyama06} 
However, we recall from the above mentioned previous comparative tests 
to the results by Singh that the thermopower is remarkably sensitive 
to details of the crystal structure. Since Usui {\em et al.}\ used 
slightly different lattice constants in their calculations this might 
explain the systematic deviations in the calculated thermopower.

\subsection{Electrical resistivity: A metal-insulator transition}

\begin{figure}[htb]
\centering
\includegraphics[width=\columnwidth,clip]{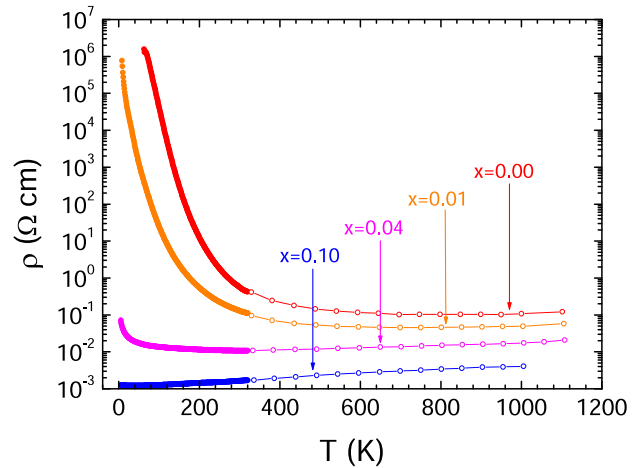}
\caption{(Color online) Temperature dependence of the resistivity of 
         $ {\rm CuRh_{1-x}Mg_xO_2} $ for $ {\rm x=0} $, $ {\rm x=0.01} $, 
         $ {\rm x=0.04} $, and $ {\rm x=0.10} $.} 
\label{fig:rho}
\end{figure}

$\rho(T)$ curves for the $ {\rm CuRh_{1-x}Mg_xO_2} $ series as given in 
Fig.~\ref{fig:rho} 
reveal the modification of the electronic groundstate induced by the 
substitution.  At 300\,K, the  values decrease by a factor of $\sim 300$ 
as $ {\rm x}$ increases from $ {\rm x=0.00} $ to $ {\rm x=0.10} $. This drop is 
even more pronounced at lower temperatures as the $\rho(T)$  curve 
exhibits a localized behavior in $ {\rm CuRhO_2} $ with 
$\rho_{\rm 100K}= 3~10^4\,\Omega~{\rm cm} $ at 100\,K. In contrast, 
metal-like behavior is observed for $ {\rm CuRh_{0.90}Mg_{0.10}O_2} $
with $\rho_{\rm 100K}= 1.2~10^{-3}\,\Omega~{\rm cm} $. It 
must be emphasized that the change of electronic state induced by 
$ {\rm Mg^{2+}} $ is progressive. The $ {\rm CuRh_{0.99}Mg_{0.01}O_2} $ 
compound still exhibits a localized behavior but with a resistivity 
decreased by three orders of magnitude at 100\,K as compared to 
$ {\rm CuRhO_2} $, whereas for $ {\rm CuRh_{0.96}Mg_{0.04}O_2} $,
$\rho$ remains almost $ T $ independent from 100\,K  to 1000\,K 
($\rho$ increasing by 5\% only in this temperature range). The $\rho(T)$ 
curve shows a re-entrant behavior only below $\sim 100$\,K, reaching 
a maximum value of $\rho = 8~10^{-1}\,\Omega~{\rm cm}$ at 5\,K. 
In fact, a closer inspection of the curves reveals that they all go 
through a minimum value at a characteristic temperature 
$ {T_{\rm min}} $ separating a $\frac{d \rho}{dT}<0$ regime below 
$ {T_{\rm min}} $ from a $\frac{d \rho}{dT}>0$ regime for 
$ {T> T_{\rm min}} $ (Fig.~\ref{fig:rho}). As shown in
Table~\ref{tab:1}, 
\begin{table}[b]
\begin{tabular}{|c|c|c|}
\hline
  x  & $ {T_{\rm min}} $(K) & $ \rho({T_{\rm min}}) {\rm (m\Omega~cm)} $ \\
\hline
0.00 &   800   &   102    \\
0.01 &   699   &    45.56 \\
0.04 &   320   &    10.74 \\
0.10 &    38   &     1.22 \\
\hline
\end{tabular}
\caption{$ {T_{\rm min}} $ and doping-dependence of the resistivity at 
         $ {T_{\rm min}} $ for several compositions of 
         $ {\rm CuRh_{1-x}Mg_{x}O_2} $.}
\label{tab:1}
\end{table} 
the $ {T_{\rm min}} $ value decreases from $ {T_{\rm min}} = 800 $\,K 
for $ {\rm CuRhO_2} $  to $ {T_{\rm min}} = 38 $\,K for 
$ {\rm CuRh_{0.90}Mg_{0.10}O_2} $.

For the pristine compound, the localizing behavior observed below 
$ T_{\rm min} $ (800\,K) is consistent with the existence of a rather 
small gap at the Fermi level obtained in the section \ref{sec:calc}. 
Even though this temperature dependence of the resistivity is similar 
to the curve measured for $ {\rm CuCrO_2} $, it is in fact closer to 
the one of $ {\rm CuCr_{0.99}Mg_{0.01}O_2} $.\cite{cucro2} However, 
the data point to a different transport mechanism. Indeed, first, 
no significant magneto-resistance was observed in contrast to all 
$ {\rm CuCr_{1-x}Mg_xO_2} $ samples showing a magneto-resistance as 
high as $ -10 $\% at 5\,K in 7\,T. Second, neither the Arrhenius law 
$\rho \propto e^{-T_0/T} $  nor the polaronic model 
$\rho \propto T e^{-T_0/T} $, nor the variable range hopping model \mbox{$\rho
\propto e^{-(T_0/T)^{\alpha}} $} with $\alpha=1/2$, $1/3$ or $1/4$, which are
broadly used in conventional three 
dimensional transition metal perovskites as $ {\rm La_{1-x}Sr_{x}Co O_3}
$,\cite{Smith08} and which were successfully applied to the two dimensional
chromium based delafossites 
$ {\rm CuCrO_2} $ and $ {\rm CuCr_{0.98}Mg_{0.02}O_2} $, 
respectively,\cite{cucro2} can convincingly fit the $\rho(T)$ data 
of $ {\rm CuRhO_2} $.

\begin{figure}[htb]
\centering
\includegraphics[width=\columnwidth,clip]{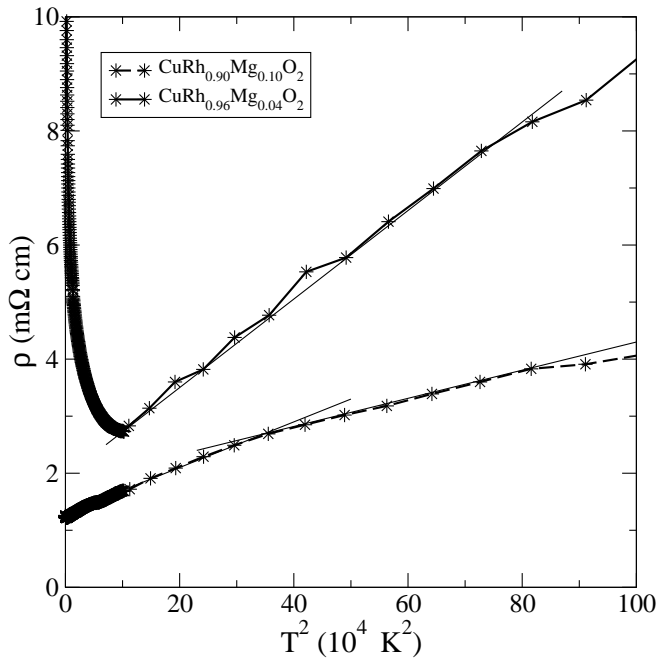}
\caption{(Color online) 
         Fermi liquid-like behavior of $ {\rm CuRh_{1-x}Mg_xO_2} $ 
         for the dopings $ {\rm x=0.04} $ and $ {\rm x=0.1} $. For
         $ {\rm CuRh_{0.96}Mg_{0.04}O_2} $, a constant contribution 
         $\rho_{cc} = 8\,{\rm m\Omega~cm}$ has been subtracted 
         for clarity. The thin full lines represent the Fermi liquid 
         fits, while the thick full lines are guides for the eyes only.}  
\label{fig:fl}
\end{figure}

This conclusion about the transport mechanism in $ {\rm CuRhO_2} $ is 
confirmed by the analysis of the $\rho(T)$ curves found for the 
Mg-substituted $ {\rm CuRhO_2} $ compounds. First, the attempts to fit 
the curves by a polaronic model fail as for $ {\rm CuRhO_2} $. Second, a low
temperature Fermi liquid behavior is found for 
$ {\rm CuRh_{0.90}Mg_{0.10}O_2}$. Third, 
the metal-like regions of the curves can be adjusted to $T^2$ 
dependences that are obeyed over a wide temperature range. As shown 
in Fig.~\ref{fig:fl} 
\begin{table}[b]
\begin{tabular}{|c|c|}
\hline
 Compound & $ A {\rm (\Omega~m~K^{-2})} $ \\
\hline
$ {\rm CuRh_{0.96}Mg_{0.04}O_2} $  & 
                   $ 7.7~10^{-11} $ \\
$ {\rm CuRh_{0.90}Mg_{0.10}O_2} $    & 
                   $ 3~10^{-11} $ \\
$ {\rm PdCoO_2} $ (from \protect Ref.\ \onlinecite{Tanaka96}) &
                   $ 4.8~10^{-13} $ \\
Thin films of $ {\rm V_2O_3} $ (from \protect Ref.\ \onlinecite{Grygiel07}) & 
                   $ 2~10^{-9} $ \\
$ {\rm La_{1-x}Sr_{x}TiO_3} $ (from \protect Ref.\ \onlinecite{Tokura93}) & 
                   $ 2 $-$ 3~10^{-11} $ \\
\hline
\end{tabular}
\caption{Fermi liquid transport parameter for several oxides.}
\label{tab:2}
\end{table} 
for $ {\rm CuRh_{0.96}Mg_{0.04}O_2} $, the $\rho \propto T^2$ regime  
holds from $\sim 300$\,K to $\sim 850$\,K. Expressing $\rho(T)$ as 
$ \rho(T)= \rho_0 + AT^2 $ as in a Fermi liquid we obtain the transport 
parameter $A$. As shown in Table~\ref{tab:2}, the values of $A$ for these rhodates are found to be located in an 
intermediate range, namely they are larger than in the $ {\rm PdCoO_2} $
delafossites,\cite{Tanaka96} but smaller than in thin films of 
$ {\rm V_2O_3} $.\cite{Grygiel07} In a fashion similar to the behavior 
observed in the titanates $ {\rm La_{1-x}Sr_{x}TiO_3}$,\cite{Tokura93} $A$
increases with decreasing  
concentration of charge carriers, while, in contrast to all the above 
systems, the $T^2$-behavior may be observed for temperatures up to 
1000\,K. Remarkably, the widely observed phonon-dominated 
$\rho \propto T$ behavior is not taking over, even at such
high temperatures. Such a result enlights the unusual transport behavior 
of these rhodates. The presence of the $ {\rm Mg^{2+}} $ scattering 
centers, up to 12\%, corresponds to concentration well below the 
percolation threshold, and no band purely based on Mg orbitals is 
expected to form. Therefore, on their own, such low Mg concentrations should
not affect the transport in the $ T $ range where this $ \rho \propto T^2 $ 
regime is observed but by hole doping the Rh-based $ 4d $ bands. Still,
inhomogeneous distribution of the Mg ions on the Rh sites might be responsible
of 
the kink observed in Fig.~\ref{fig:fl} for 
$ {\rm CuRh_{0.90}Mg_{0.10}O_2} $.

Finally, the resistivity values for $ {\rm CuRh_{1-x}Mg_xO_2} $ are 
rather comparable to those reported in Ref.\ \onlinecite{shibasaki06} 
($T\leq~\!\!300$\,K) or in Ref.\ \onlinecite{kuriyama06} ($T \geq 400$\,K). 
The decrease of $\rho$ induced by the $ {\rm Mg^{2+}} $ substitution 
in the present samples strongly suggest that ``hole'' charge carriers 
are created  according to the formula 
$ {\rm CuRh_{1-2x}^{3+}Rh_x^{4+}Mg_x^{2+}O_2} $. It must also be
added that since no other $ {\rm CuMO_2} $ delafossites exhibits such 
metal-like behavior down to very low $ T $, the role of the Cu channel 
to the electronic transport can hardly be invoked as is also confirmed 
by the electronic structure calculations.

\subsection{Thermoelectric power and power factors} 

\begin{figure}[htb]
\centering
\includegraphics*[width=\columnwidth,clip]{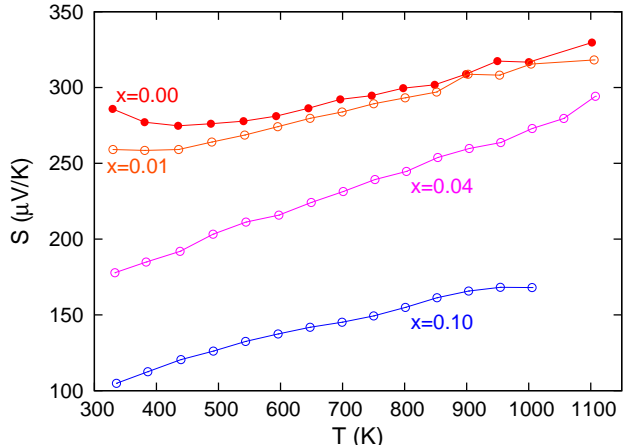}
\caption{(Color online) Temperature dependence of the thermopower of 
         $ {\rm CuRh_{1-x}Mg_xO_2} $ for $ {\rm x=0} $, $ {\rm x=0.01} $, 
         $ {\rm x=0.04} $, and $ {\rm x=0.10} $. Lines are guide for the 
         eyes, only.} 
\label{fig:tep}
\end{figure}

Although the $\rho$ values of our $ {\rm CuRh_{1-x}Mg_xO_2} $ series are
comparable to those already reported,\cite{kuriyama06} the Seebeck
coefficients values for $ {\rm CuRhO_2} $ (Table~\ref{tab:3} and
Fig.~\ref{fig:tep}) appear to be different. 
\begin{table}[b]
\begin{tabular}{|c|c|c|}
\hline
x                     & $ T $(K) & $ S {\rm (\mu V~K^{-1}) } $ \\
\hline
0                                                        & 330 & 280  \\
0 (from \protect Ref.\ \onlinecite{shibasaki06})         & 300 & 130  \\
0.10                                                     & 450 & 120  \\
0.10 (from \protect Ref.\ \onlinecite{kuriyama06})       & 450 & 165  \\
0.05-0.20 (from \protect Ref.\ \onlinecite{shibasaki06}) & 450 &  65  \\
\hline
\end{tabular}
\caption{Typical values of the thermopower of $ {\rm CuRh_{1-x}Mg_xO_2} $.}
\label{tab:3}
\end{table} 
When compared to the study in which the samples were calcined at lower 
$ T $, \cite{shibasaki06} the present $S$ values are found to be always 
much larger, as documented in Table~\ref{tab:3}. Besides, our data for
the substituted compounds showing an $S$ decrease as x increases cannot be 
reconciled with the x-independent $S_{\rm 300K} $ for all substituted 
compounds of Ref.\ \onlinecite{shibasaki06}. 

For the most metallic sample, the $S$ values are always increasing with 
$T$ as shown for $ {\rm CuRh_{0.90}Mg_{0.10}O_2} $ in Fig.~\ref{fig:tep}. 
In contrast, the localizing behavior of $ {\rm CuRhO_2} $ is reflected 
by the $S(T)$ curves showing an upturn towards high $S$ values as $T$ 
decreases below $\sim$ 450\,K. For all $ {\rm x > 0.01}$,  or for $T > 450 $\,K
(for x = 0.00 and x = 0.01) the $S(T)$ curves exhibit an almost 
$S \propto T$ regime as shown by the lines drawn on the $S(T)$ curves 
in Fig.~\ref{fig:tep}. Such a $T$ dependence is 
characteristic of a metallic behavior. As previously 
reported for $ {\rm CuRh_{0.90}Mg_{0.10}O_2} $, this behavior leads to 
rather large positive values, larger than $300\,{\rm \mu V~K^{-1}} $ 
at 1000\,K for $ {\rm x = 0.00} $ and $ {\rm x = 0.01} $. The decrease 
of $S$ as x increases together with the positive sign of $S$ is consistent 
with an increase of the hole ($ {\rm Rh^{4+}} $) fraction induced by charge 
compensation created by the $ {\rm Mg^{2+}} $ for $ {\rm Rh^{3+}} $ 
substitution.

Remarkably, all these trends are in good qualitative agreement with the 
theoretical findings presented in Sec.\ \ref{sec:calc}. Yet, the 
experimentally obtained values at 1000\,K are somewhat smaller than the 
theoretically predicted ones. One is therefore tempted to analyze the 
above experimental data within a completely different approach, namely, 
the temperature independent correlation functions ratio approximation
(TICR).\cite{fresard02}  
In this approach, the thermopower arising from the Kubo formalism
\begin{equation}
S(T) = \frac{1}{eT} \frac{\langle j_E j_n \rangle - \mu \langle j_n j_n\rangle}
{\langle j_n j_n \rangle} 
\label{eq:kubo}
\end{equation}
is approximated by assuming that the energy current-particle current 
correlation function and the particle current autocorrelation function 
share the same temperature dependence. Thus their ratio should result 
in a hyperbolic offset of strength $E_0$ of the thermopower as 
\begin{equation}
S(T) = \frac{1}{eT} ( E_0 - \mu(T) ) \;.
\label{eq:est}
\end{equation}
At high temperatures the temperature dependence of the thermopower is 
governed by the one of the chemical potential which follows from 
\begin{equation}
n(T) = \int \mathrm{d} \epsilon f(E - \mu) \rho(E) 
                        \;.
\label{eq:mu}
\end{equation}
Here, $ f(E) $ again denotes the Fermi function and $ \rho(E) $ is the 
density of states. In the present context, the latter is taken from the 
electronic structure calculations (Fig.~\ref{fig4}). Yet, within the 
present model there is still room for improvement especially at high 
temperatures. This goes mainly along two different directions. One way 
to improve on the TICR as covered by (\ref{eq:est})/(\ref{eq:mu}) would 
be to use a modified density of states resulting from the GGA result 
by a rigid energetical upshift of the conduction bands. However, for 
reasonable values of the latter, it turned out that the thermopower 
shows little sensitivity to the actual optical band gap. From this, 
we can furthermore conclude that the behavior of the thermopower is 
dominated by the holes in the valence bands. 

Another direction is provided by the observation that, in many cases, 
the thermopower tends to loose its temperature dependence above room 
temperature, see e.~g. Ref.\ \onlinecite{Limelette06}. 
With this motivation, the TICR expression (\ref{eq:est}) was extended 
by adding a temperature independent contribution $S_0$ to $S(T)$.  
\begin{figure}[b]
\centering
\includegraphics*[width=\columnwidth,clip]{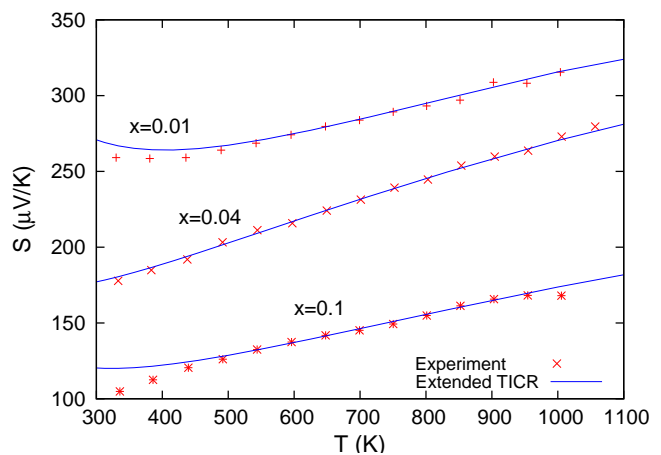}
\caption{(Color online) Comparison of the different theoretical models, 
         including the temperature independent contribution $S_0$, 
         (lines) with our experimental values (symbols) of the 
         temperature dependence of the thermopower of 
         $ {\rm CuRh_{1-x}Mg_xO_2}$ for ${\rm x=0.01}$, ${\rm x=0.04}$, 
         and ${\rm x=0.10}$} 
\label{fig:mudeT}
\end{figure}
Such a contribution is
often attributed to localized degenerate states. Here they may derive from
the pockets on the Fermi surface centered around the H and A points that are
characterized by small Fermi velocities (see Fig.~\ref{fig3}). As a 
consequence, within the extended TICR the thermopower is determined by two 
parameters, namely, $ E_0 $ and $ S_0 $, which, together with the GGA DOS, 
can be used to fit the experimental 
data. The result is shown in Fig.~\ref{fig:mudeT} for all three doped 
samples. Regarding  $E_0$, the values range from $40$~meV to $90$~meV, 
thereby being quite similar to the ones reported for electron-doped 
manganites. In these materials a large density of states at the bottom of 
the band results in low degeneracy
temperatures and large negative thermopower.\cite{fresard02} For the title
compounds, the density of states at the top of the valence band is large,
resulting in the large positive thermopower, together with low degeneracy
temperature. The $S_0$ values
are quite large ($40 \mu V K^{-1} \leq S_0 \leq 100 \mu V K^{-1}$) and appear
to be comparable to the ones reported for various layered colbatates,  in
which the here considered $\rm{RhO_2}$ layers are replaced by isostructural
$\rm{CoO_2}$ layers.\cite{Limelette06} Additionally, the fact that the
thermopower data for all three doped samples could be reproduced using the
same DOS for both the extended TICR model and the GGA calculations is quite
remarkable. This gives a strong support to use the rigid band model.

\begin{figure}[htb]
\centering
\includegraphics*[width=\columnwidth,clip]{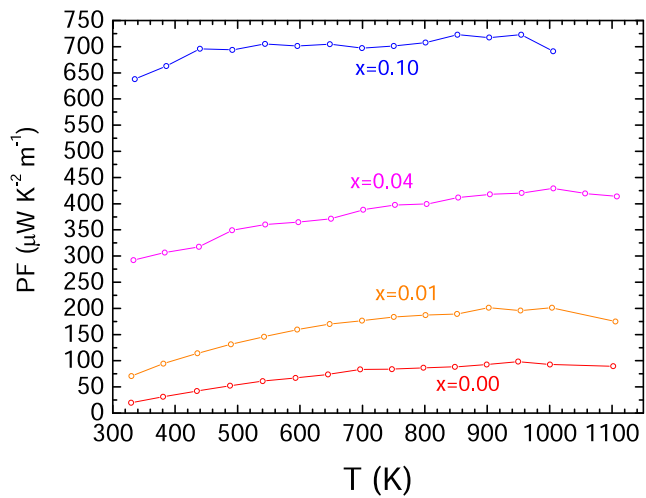}
\caption{(Color online) Temperature dependence of the power factor of 
         $ {\rm CuRh_{1-x}Mg_xO_2} $ for $ {\rm x=0} $, $ {\rm x=0.01} $, 
         $ {\rm x=0.04} $, and $ {\rm x=0.10} $. Lines are guide for the 
         eyes, only.} 
\label{fig:pf}
\end{figure}

In order to check the thermoelectric performance, we now address the power
factor PF ($ {\rm PF} =\frac{S^2}{\rho} $). Combining their $T$ dependences, 
$S \propto T$ and $\rho \propto T^2$ at sufficiently high temperature, 
this leads to remarkable $ T $-independent values of the PF as shown in 
Fig.~\ref{fig:pf}. 
The best values are observed for the most substituted samples exhibiting 
a value of the $ {\rm PF= 7~10^{-4} W K^{-2} m^{-1}} $. As shown 
in Fig.~\ref{fig:pf} and in good qualitative agreement with the 
calculations presented in Sec.\ \ref{sec:calc}, the values are found to 
increase as x increases in $ {\rm CuRh_{1-x}Mg_xO_2} $ showing that the 
induced relative $\rho$ decrease is more than compensating the $S$ decrease.

\section{Conclusion}

In summary, Mg-doped $ {\rm CuRhO_2} $ has been investigated by means 
of electronic structure calculations, structural characterization, and 
transport measurements. The electronic structure calculations clearly
indicate that the transport is dominated by the Rh $ 4d $ bands. Structural
data demonstrate that the solubility limit of Mg in $ {\rm CuRhO_2} $ is as
high as 12\%, provided the samples are prepared at temperatures above $\sim$
$1000^{\circ}$C, in which case Rh is indeed substituted by Mg. This
substitution results in a peculiar hole doping of the rather narrow Rh $ 4d $
bands, shown by a $T^2$ dependence of the resistivity, a $T$ dependence of the
thermopower, and a quite large nearly $T$ independent power factor, up to
temperatures as high as 1000\,K. Regarding the thermopower, good qualitative
agreement between the theoretical prediction arising from the GGA+Boltzmann
approach and experimental data is obtained. Yet, an
additional purely entropic contribution needs to be invoked when treating the
GGA results in the TICR framework for
quantitative explanation, in
a similar fashion as what was found in cobaltites 
with isostructural $ {\rm CoO_2}$ layers.\cite{Limelette06} Thermal
conductivity measurements performed on dense samples would be necessary to
measure the figure of merit. Work along this line is in progress. 

\section{Acknowledgments}

We gratefully acknowledge many useful discussions with T.\ Kopp and 
A.\ Reller. V.E.\ is especially grateful to D.\ Singh for very fruitful
discussions during the implementation of the transport properties into the ASW
package. This work was supported by the ANR through NEWTOM as well as 
by the Deutsche Forschungsgemeinschaft through SFB 484 (V.E.) and 
the Research Unit 960 ``Quantum Phase Transitions'' (S.K.).


\begin{thebibliography}{}
\bibitem{kuriyamarem}
This value ($130\,{\rm \mu V K^{-1}} $) is extrapolated from the high-temperature measurements reported
in Ref.\ \onlinecite{kuriyama06}.
\end{thebibliography}
\end{document}